\documentclass[aps,prd,onecolumn,groupedaddress,showpacs,nofootinbib,amssymb]{revtex4}
\usepackage[dvips]{graphicx}
\usepackage{amssymb}
\usepackage{amsmath}
\usepackage{graphicx}
\usepackage{amsfonts}
\usepackage{bm}

\begin{document}

\title{Mimetic $F(R)$ inflation confronted with Planck and BICEP2/Keck Array data}

\author{S.D. Odintsov$^{1,2}$, V.K. Oikonomou$^{3,4}$}
\affiliation{$^1$Instituci\`{o} Catalana de Recerca i Estudis Avan\c{c}ats
(ICREA),
Barcelona, Spain \\
$^2$Institut de Ciencies de l'Espai (CSIC-IEEC), Campus UAB,
Campus UAB, Carrer de Can Magrans, s/n 08193 Cerdanyola del Valles,
Barcelona, Spain\\
$^{3)}$ Tomsk State Pedagogical University, 634061 Tomsk \\
$^{4)}$  Laboratory for Theoretical Cosmology, Tomsk State University of Control Systems
and Radioelectronics (TUSUR), 634050 Tomsk, Russia
}

\begin{abstract}
In this paper we demonstrate that in the context of mimetic $F(R)$ gravity with Lagrange multiplier, it is possible to realize cosmologies which are compatible with the recent BICEP2/Keck Array data. We provide some characteristic examples for which the predicted scalar to tensor ratio can be quite smaller in comparison to the upper limit imposed by the BICEP2/Keck Array observations.
\end{abstract}

\pacs{95.35.+d, 98.80.-k, 98.80.Cq, 95.36.+x}

\maketitle

\section{Introduction}

The late-time acceleration of our Universe was one of the most striking observations made the last twenty years \cite{latetimeobse}. Ever since a lot of effort was made towards the consistent theoretical explanation of this late-time acceleration. The modified theories of gravity, and specifically $F(R)$ gravity (for reviews see \cite{reviews1,bam1}), have a prominent role for the theoretical modelling of this late-time acceleration and in general for the description of our Universe's cosmological evolution. For some important reviews with respect to $F(R)$ gravity, we refer the reader to \cite{reviews1,bam1} and also for some insightful early studies on the subject, see \cite{barrowearly}. But apart from the late-time acceleration modelling, the $F(R)$ gravities also describes early-time acceleration, which is known as inflation \cite{inflrev,bam1}, but also bouncing cosmology \cite{bounce}. In fact in some cases it is possible to provide a unified description of early and late-time acceleration by using a single model of $F(R)$ gravity, as was done in \cite{sergeinojiri2003}, see also \cite{bam1}.

The observed composition of our Universe consists of three components, the dark energy which contributes the larger part of our Universe's density $\Omega_{DE}\sim 68.3\%$, the ordinary baryonic, or luminous matter, with density $\Omega_m\sim4.9\%$, and the so-called dark matter, with density $\Omega_{DM}\sim 26.8\%$. Dark energy and dark matter are still mysterious, and there exist many potential descriptions which could successfully describe both, but need to be observationally or experimentally verified. Dark matter is believed to be of particle nature, and there exist many theoretical models which describe dark matter as particle, see for example \cite{darkmatter,shafi}. Apart from the particle description of dark matter, there also exist other mechanisms that can mimic dark matter, such as the recently proposed mimetic dark matter mechanism proposed by Chamseddine and Mukhanov \cite{mukhanov1}. The mimetic dark matter mechanism is a modification of general relativity in which the conformal degrees of freedom of the physical metric of spacetime is taken into account. For some further attempts to generalize and study cosmological applications of mimetic modified gravity theory, see \cite{mukhanov2,Golovnev}, and for various extensions and studies on the cosmology of mimetic modified gravity, see \cite{mimetic1}. As we mentioned, the mimetic dark matter formalism employs a covariant approach which isolates the conformal degree of freedom of the physical metric in ordinary Einstein Hilbert gravity. Therefore, the physical metric is expressed in terms of an auxiliary scalar degree of freedom $\phi$, with it's first derivatives appearing in the gravitational action, and in effect, the conformal degrees of freedom become dynamical. Eventually, the resulting dynamical conformal degree of freedom mimics cold dark matter, even in the absence of an ordinary cold dark matter component in the energy momentum tensor. The mimetic dark matter formalism was extended in the context of $F(R)$ gravity in Ref. \cite{NO2}, with the theoretical framework also containing a Lagrange multiplier and a scalar potential. In the resulting mimetic $F(R)$ gravity, the unified description of early and late-time acceleration can easily be achieved, even in the vacuum $F(R)$ gravity case. Actually, in Refs. \cite{mimeletter,mimeticbig} we used the mimetic $F(R)$ gravity framework, to investigate which models can be compatible with the recent Planck data \cite{planck}, and as we demonstrated, concordance with observations can easily be achieved. In fact, by using an arbitrary $F(R)$ gravity, we achieved compatibility by suitably choosing the mimetic scalar potential and the Lagrange multiplier. It is exactly the presence of the mimetic potential and of the Lagrange multiplier that allows us to have compatibility with observational data, regardless the choice of the $F(R)$ gravity. Actually, in some cases, the predictions were particularly appealing, especially with regards to the scalar to tensor ratio $r$, which was much smaller than the upper limit allowed by the Planck data. In this paper we shall exploit this feature to show that the mimetic $F(R)$ gravity can produce models which are compatible firstly with the Planck data \cite{planck} and secondly to show that our results are compatible with the recently released BICEP2/Keck Array data \cite{bicep2}. The results of the BICEP2/Keck Array impose quite stringent constraints on the scalar-to-tensor ratio and our aim in this paper is to point out that even in this case, the mimetic $F(R)$ gravity can produce compatible to the observational data results, and this study further support the utility of our first two works on the subject \cite{mimeletter,mimeticbig}. Hence in the present work the novelty is that we also take into account the recently released observational data, and we show that the mimetic $F(R)$ models can be compatible with even more stringent constraints, with regards to the Planck data, which are considered to be accurate to the bounds which impose on the observational indices.

Recently, it was reported by the BICEP2/Keck Array collaboration \cite{bicep2} that the upper limit of the scalar to tensor ratio is further restricted to the value $r<0.07$. The purpose of this paper is to briefly demonstrate that in the context of mimetic $F(R)$ gravity, it possible to achieve values for the scalar to tensor ratio $r$, well below the upper limit predicted by the BICEP2/Keck Array collaboration. The formalism of mimetic $F(R)$ gravity includes the use of a mimetic scalar potential of the auxiliary scalar field $\phi$, which we denote $V(\phi)$, and also of a Lagrange multiplier scalar function of the auxiliary field too, which we denote $\lambda (\phi)$. As was also shown in \cite{mimeletter,mimeticbig}, by using the Lagrange multiplier formalism developed in \cite{CMO}, we can have a new reconstruction technique which effectively enables us to realize an arbitrary cosmological evolution, by suitably choosing the $F(R)$ gravity, the Lagrange multiplier and the mimetic potential. Hence, the freedom provided by the Lagrange multiplier and the scalar mimetic potential, can effectively enable us to use a cosmological evolution compatible with the Planck \cite{planck} and BICEP2/Keck Array \cite{bicep2} observational data.

This paper is organized as follows: In section II, we briefly review all the important features of the mimetic $F(R)$ gravity with Lagrange multiplier and mimetic potential. In section III we apply the formalism to produce two cosmological models which are compatible with the recent BICEP2/Keck Array data. Particularly, we are interested in the values of the spectral index of primordial curvature perturbations and the scalar to tensor ratio. By using a simple $F(R)$ gravity, we explicitly demonstrate that in both cases, compatibility can be achieved for quite general values of the free parameters of the theory. Finally, the conclusions follow in the end of the paper.

\section{Lagrange-multiplier Mimetic $F(R)$ Gravity Formalism}

In the context of mimetic $F(R)$ gravity with Lagrange multiplier, the conformal symmetry is considered to be a non-violated internal degree of freedom \cite{mukhanov1}. As was firstly considered in \cite{mukhanov1}, the physical metric, which we denote by $g_{\mu \nu}$, can be written in terms of the auxiliary scalar $\phi$ and in terms of an auxiliary metric $\hat{g}_{\mu \nu}$, in the following way,
\begin{equation}\label{metrpar}
g_{\mu \nu}=-\hat{g}^{\mu \nu}\partial_{\rho}\phi \partial_{\sigma}\phi
\hat{g}_{\mu \nu}\, .
\end{equation}
A consequence of Eq. (\ref{metrpar}) is the following constraint,
\begin{equation}\label{impl1}
g^{\mu \nu}(\hat{g}_{\mu \nu},\phi)\partial_{\mu}\phi\partial_{\nu}\phi=-1\,
.
\end{equation}
By performing a Weyl transformation of Eq. (\ref{metrpar}), we can easily conclude that it remains invariant, and in effect the auxiliary metric does not appear in the final gravitational action and the corresponding equations of motion. In the following we shall assume that the physical metric $g_{\mu \nu}$, is a flat Friedman-Robertson-Walker (FRW) metric with line element,
\begin{equation}\label{frw}
ds^2 = - dt^2 + a(t)^2 \sum_{i=1,2,3}
\left(dx^i\right)^2\, ,
\end{equation}
where $a(t)$ stands for the scale factor. For the flat FRW metric (\ref{frw}), the Ricci scalar is,
\begin{equation}\label{ricciscalarnnew}
R=6\left (\dot{H}+2H^2 \right )\, ,
\end{equation}
and furthermore, we shall assume that the scalar field $\phi$ depends only on the cosmic time, that is $\phi=\phi (t)$. The Jordan frame mimetic vacuum $F(R)$ gravity with Lagrange multiplier theory, has the following Jordan frame action\cite{NO2},
\begin{equation}\label{actionmimeticfraction}
S=\int \mathrm{d}x^4\sqrt{-g}\left ( F\left(R(g_{\mu \nu})\right
)-V(\phi)+\lambda \left(g^{\mu \nu}\partial_{\mu}\phi\partial_{\nu}\phi
+1\right)\right )\, .
\end{equation}
Note that, the mimetic $F(R)$ gravity formalism can accordingly be extended in other modified gravity theories, with Lagrangian $L= \sqrt(-g)\left(F(R, R^{\mu \nu}R_{\mu \nu} , R_{ \mu \nu \alpha \beta} R^{ \mu \nu \alpha \beta} \right))-V(\phi )+\lambda
\partial_{\mu}\phi \partial_{\nu}\phi$, or for the non-local gravity case, with the following Lagrangian $L=\sqrt(-g)\left(F(R,R\square^m R,\square^d R)\right)-V(\phi)\lambda
\partial_{\mu}\phi \partial_{\nu}\phi $, e.t.c., where the parameters $m$ and $d$ can either be positive or negative. By varying the gravitational action (\ref{actionmimeticfraction}) with respect to the metric $g_{\mu \nu}$, and with respect to the mimetic scalar field, we obtain the following equations of motion (see \cite{mimeticbig} for more details on this),
\begin{equation}\label{enm1}
-F(R)+6(\dot{H}+H^2)F'(R)-6H\frac{\mathrm{d}F'(R)}{\mathrm{d}t}-\lambda
(\dot{\phi}^2+1)+V(\phi)=0\, ,
\end{equation}
\begin{equation}\label{enm2}
F(R)-2(\dot{H}+3H^2)+2\frac{\mathrm{d}^2F'(R)}{\mathrm{d}t^2}+4H\frac{\mathrm{d}F'(R)}{\mathrm{d}t}-\lambda (\dot{\phi}^2-1)-V(\phi)=0\, ,
\end{equation}
\begin{equation}\label{enm3}
2\frac{\mathrm{d}(\lambda \dot{\phi})}{\mathrm{d}t}+6H\lambda
\dot{\phi}-V'(\phi)=0\, ,
\end{equation}
\begin{equation}\label{enm4}
\dot{\phi}^2-1=0\, ,
\end{equation}
where the ``dot'' indicates differentiation with respect to the cosmic time, and in this case, the prime denotes differentiation with respect to the Ricci scalar $R$ in Eqs. (\ref{enm1}) and (\ref{enm2}), while in Eq. (\ref{enm3}), it indicates differentiation with respect to the scalar field $\phi$. From Eq. (\ref{enm3}), it easily follows that the auxiliary scalar can be identified with the cosmic time $t$, and therefore we can rewrite Eq. (\ref{enm2}), in the following way,
\begin{equation}\label{sone}
F(R)-2(\dot{H}+3H^2)+2\frac{\mathrm{d}^2F'(R)}{\mathrm{d}t^2}+4H\frac{\mathrm{d}F'(R)}{\mathrm{d}t}-V(t)=0\, .
\end{equation}
From Eq. (\ref{sone}), we have an analytic expression that yields the mimetic scalar potential as a function of cosmic time and expressed in terms of the $F(R)$ gravity function and the Hubble rate, in the following way,
\begin{equation}\label{scalarpot}
V(\phi=t)=2\frac{\mathrm{d}^2F'(R)}{\mathrm{d}t^2}+4H\frac{\mathrm{d}F'(R)}{
\mathrm{d}t}+F(R)-2(\dot{H}+3H^2)\, .
\end{equation}
Upon combining of Eqs. (\ref{scalarpot}) and (\ref{enm1}), we obtain the Lagrange multiplier function too, expressed in terms of $F(R)$ gravity function and the Hubble rate, as follows,
\begin{equation}\label{lagrange}
\lambda (t)=-3 H \frac{\mathrm{d}F'(R)}{\mathrm{d}t}+3
(\dot{H}+H^2)-\frac{1}{2}F(R)\, .
\end{equation}
Finally, by combining Eqs. (\ref{scalarpot}) and (\ref{lagrange}), we easily obtain the following differential equation,
\begin{equation}\label{diffeqnnewaddition}
2\frac{\mathrm{d}^2F'(R)}{\mathrm{d}t^2}+4H\frac{\mathrm{d}F'(R)}{\mathrm{d}
t}- 6 H \frac{\mathrm{d}F'(R)}{\mathrm{d}t}-2 (\dot{H}+3
H^2)+6(\dot{H}+H^2)\frac{\mathrm{d}F'(R)}{\mathrm{d}t}-2 \lambda
(t)-V(t)=0\, \end{equation}
which will prove to be very useful in the following sections. By observing Eqs. (\ref{scalarpot}), (\ref{lagrange}) and (\ref{diffeqnnewaddition}), we can see that we have a very conceptually simple reconstruction method, in which we can realize any arbitrary cosmology, for any $F(R)$  gravity, by choosing the potential as in Eq. (\ref{scalarpot}), and the Lagrange multiplier as in Eq. (\ref{lagrange}). In the following two sections we shall extensively use this reconstruction method, in order to realize two cosmologies which are compatible with the recent BICEP2/Keck Array observational data. We need to note that the mimetic $F(R)$ gravity formalism enables us to unify acceleration of early-time or late-time universe with dark matter, due to the mimetic constraint, in a purely geometric context (see for example \cite{NO2}).

\section{Two Cosmological Models of Mimetic $F(R)$ Gravity Compatible with BICEP2/Keck Array Data}

In this section the focus is on demonstrating that two simple mimetic $F(R)$ models can be compatible with the recent BICEP2/Keck Array data. To this end we shall use a simple $F(R)$ model, and by finding the corresponding mimetic potential and Lagrange multiplier function, we shall show that the two simple models are compatible with the BICEP2/Keck Array data, focusing on the value of the scalar to tensor ratio. The $F(R)$ gravity model we shall use is the following,
\begin{equation}\label{frgravity}
F(R)=R^{1+\epsilon}\, ,
\end{equation}
where the parameters $d$ and $f$ are assumed to be arbitrary positive real numbers. Note that in the context of usual $R^{1+\epsilon}$, the parameter $\epsilon$ is strongly constrained \cite{barrowearly}. In order to calculate the spectral index of primordial curvature perturbations and the scalar to tensor ratio, we shall need the slow-roll indices calculated in the Jordan frame, so we employ the formalism developed in Ref. \cite{sergeislowroll} (see also \cite{muk}). According to Ref. \cite{sergeislowroll}, the Jordan frame $F(R)$ gravity can be treated as a perfect fluid and therefore the slow-roll indices and the corresponding observational indices can be expressed in terms of the Hubble rate and the higher derivatives of the Hubble rate.  It is more convenient to calculate the slow-roll parameters in terms of the $e$-folding number $N$, instead of the cosmic time $t$, so we transform all the relevant expressions by using the following transformations,
\begin{equation}\label{transfefold}
\frac{\mathrm{d}}{\mathrm{d}t}=H(N)\frac{\mathrm{d}}{\mathrm{d}N},{\,}{\,}{\
,}
\frac{\mathrm{d}^2}{\mathrm{d}t^2}=H^2(N)\frac{\mathrm{d}^2}{\mathrm{d}N^2}+
H(N)\frac{\mathrm{d}H}{\mathrm{d}N}\frac{\mathrm{d}}{\mathrm{d}N}\, .
\end{equation}
Then by following Ref. \cite{sergeislowroll}, the corresponding $N$-dependent slow-roll indices can be expressed as functions of $N$ as follows,
\begin{align}\label{hubbleslowrollnfolding}
&\epsilon=-\frac{H(N)}{4 H'(N)}\left(\frac{\frac{H''(N) }{H(N)}+6\frac{H'(N)
}{H(N)}+\left(\frac{H'(N)}{H(N)}\right)^2}{3+\frac{H'(N)}{H(N)}}\right)^2
\, ,\\ \notag &
\eta=-\frac{\left(9\frac{H'(N)}{H(N)}+3\frac{H''(N)}{H(N)}+\frac{1}{2}\left(
\frac{H'(N)}{H(N)}\right)^2-\frac{1}{2}\left(
\frac{H''(N)}{H'(N)}\right)^2+3
\frac{H''(N)}{H'(N)}+\frac{H'''(N)}{H'(N)}\right)}{2\left(3+\frac{H'(N)}{H(N
)}\right)}\, ,
\end{align}
where in this case, the prime indicates that expressions are differentiated with respect to the $e$-folding number $N$. By using the formalism of Ref. \cite{sergeislowroll}, the observational indices can be expressed as follows,
\begin{equation}\label{indexspectrscratio}
n_s\simeq 1-6 \epsilon +2\eta,\, \, \, r=16\epsilon \, ,
\end{equation}
where $n_s$ and $r$ are the spectral index of primordial curvature perturbations and the scalar to tensor ratio respectively. Our aim then is to find the analytic expressions of the mimetic potential and of the Lagrange multiplier function, which can produce cosmologies compatible with the Planck and BICEP2/Keck Array data. We remind that the recent Planck data indicate that \cite{planck}
\begin{equation}\label{constraintedvalues}
n_s=0.9644\pm 0.0049\, , \quad r<0.10\, ,
\end{equation}
and also the BICEP2/Keck Array further constrain the upper limit of the scalar to tensor ratio $r$, as follows,
\begin{equation}\label{scalartotensorbicep2}
r<0.07\, ,
\end{equation}
at $95\%$ confidence level. In the following two subsections we demonstrate that two simple models can be compatible with the observational constraints (\ref{constraintedvalues}) and (\ref{scalartotensorbicep2}).

\subsection{Cosmological Model I}

One simple cosmological model with quite interesting phenomenology, is described by the following Hubble rate,
\begin{equation}\label{hub2}
H(N)=\left(-G_0\text{  }e^{\alpha N }+G_1\right)^b\, ,
\end{equation}
with $G_0,G_1$ and $b$ being arbitrary parameters. By combining Eqs. (\ref{hub2}) and (\ref{hubbleslowrollnfolding}), we can obtain the slow-roll parameter $\epsilon$ which is equal to,
\begin{align}\label{hubslowroll2}
& \epsilon=-\frac{b e^{\alpha N} G_0 \alpha  \left(G_1 (6+\alpha )-2 e^{\alpha
N} G_0 (3+b \alpha )\right)^2}{4 \mathcal{F}(N)}
\end{align}
where for notational simplicity, we introduced the function $\mathcal{F}(N)$, which is defined to be equal to,
\begin{equation}\label{hfghfhgfghdf}
\mathcal{F}(N)=\left(e^{\alpha N} G_0-G_1\right) \left(-3 G_1+e^{\alpha N} G_0
(3+b \alpha )\right)^2\, .
\end{equation}
Accordingly, the slow-roll parameter $\eta$ for the Hubble rate (\ref{hub2}) reads,
\begin{equation}\label{edggs}
\eta =-\frac{\alpha  \left(8 b^2 e^{2\alpha N} G_0^2 \alpha +G_1 \left(2
e^{\alpha N} G_0 (-3+\alpha )+G_1 (6+\alpha )\right)+2 b e^{\alpha N} G_0
\left(12 e^{\alpha N} G_0-G_1 (12+5 \alpha )\right)\right)}{4 \left(e^{\alpha
N} G_0-G_1\right) \left(-3 G_1+e^{\alpha N} G_0 (3+b \alpha )\right)}
\, .
\end{equation}
By using the expressions (\ref{hubslowroll2}) and (\ref{edggs}) for the slow-roll parameters, we may easily obtain the spectral index of primordial curvature perturbation $n_s$ and the scalar to tensor ratio $r$, with $n_s$ being finally equal to,
\begin{align}\label{scalarpertandsctotenso}
& n_s=\frac{2 \left(e^N\right)^{3 \alpha } G_0^3 (3+b \alpha )^2 (1+2 b \alpha
)+3 G_1^3 \left(-6+6 \alpha +\alpha ^2\right)}{2 \mathcal{F}(N)}
+\frac{e^{\alpha N} G_0 G_1^2 \left(54+12 (-3+4 b) \alpha +3 \alpha ^2+2 b
\alpha ^3\right)}{2 \mathcal{F}(N)}\\ \notag &
-\frac{2 e^{2\alpha N} G_0^2 G_1 \left(27+(-9+48 b) \alpha +\left(3+13
b^2\right) \alpha ^2+b (1+b) \alpha ^3\right)}{2 \mathcal{F}(N)}
\, ,
\end{align}
while $r$ is equal to,
\begin{equation}\label{thodorakis}
r=-\frac{4 b e^{\alpha N} G_0 \alpha  \left(G_1 (6+\alpha )-2 e^{\alpha N} G_0
(3+b \alpha )\right)^2}{\mathcal{F}(N)}\, .
\end{equation}
The observational indices (\ref{scalarpertandsctotenso}) and (\ref{thodorakis}) can be compatible with the observational data for various values of the free parameters $G_0$, $G_1$, $\alpha$ and $b$, but a convenient choice is the following,
\begin{equation}\label{parmchoice12}
G_0=0.085,\,\,\, G_1=11,\,\,\,\alpha=0.03,\,\,\, b=1\, .
\end{equation}
By using the values (\ref{parmchoice12}) for the free parameters, the observational indices take the following values,
\begin{equation}\label{indnewparadigm12}
n_s\simeq 0.966577, \,\,\, r=0.0173982\, ,
\end{equation}
and by comparing the values (\ref{indnewparadigm12}) to the Planck data constraints of Eq. (\ref{constraintedvalues}) and also to the recent BICEP2/Keck Array data (\ref{scalartotensorbicep2}), we can easily see that the values (\ref{indnewparadigm12}) are compatible to both Planck and BICEP2/Keck Array constraints. Let us now investigate which mimetic potential and Lagrange multiplier can generate the cosmological evolution (\ref{hub2}), for the $F(R)$ gravity chosen as in Eq. (\ref{frgravity}). Introducing the function $G(N)$, defined to be, $G(N)=H(N)^2$
(see also Ref. \cite{sergeirecon}), we can express the Ricci scalar in terms of the function $G(N)$, as follows,
\begin{equation}\label{cnc}
R(N)=3G'(N)+12G(N)\, .
\end{equation}
Then, by using Eqs. (\ref{frgravity}), (\ref{hub2}), (\ref{cnc}), and
(\ref{scalarpot}), we can find the analytic form of the potential $V(N)$ which generates the cosmology (\ref{hub2}), which for simplicity we presented in the Appendix. We can find an approximate form of the potential, by expanding it in powers of the parameter $\epsilon$. Note that when $\epsilon$ is small, then the $F(R)$ gravity of Eq. (\ref{frgravity}) describes a small modification of general relativity. Performing the $\epsilon$-expansion, the mimetic potential reads,
\begin{align}\label{podnegsa}
& V(N)= \left(3 \mathcal{S}(N)^{2 b} \left(4+\mathcal{S}(N)^{2 b}\right)-2 \left(3 \mathcal{S}(N)^{2 b}-b \left(e^N\right)^{\alpha } G_0 \mathcal{S}(N)^{-1+2 b} \alpha \right)\right)\\ \notag &
+\Big{(} \frac{4}{3 \left(4+\mathcal{S}(N)^{2 b}\right)}+2 \Big{(}(\frac{4 b \left(e^N\right)^{\alpha } G_0 \mathcal{S}(N)^{2 b} \left(2+\mathcal{S}(N)^{2 b}\right) \alpha }{\left(\left(e^N\right)^{\alpha } G_0-G_1\right) \left(4+\mathcal{S}(N)^{2 b}\right)} \\ \notag &
-\frac{4 b \left(e^N\right)^{\alpha } G_0 \mathcal{S}(N)^{2 b} \left(8 G_1-4 b \left(e^N\right)^{\alpha } G_0 \mathcal{S}(N)^{2 b}+6 G_1 \mathcal{S}(N)^{2 b}+G_1 \mathcal{S}(N)^{4 b}\right) \alpha ^2}{\left(\left(e^N\right)^{\alpha } G_0-G_1\right)^2 \left(4+\mathcal{S}(N)^{2 b}\right)^2}\Big{)}\\ \notag &
+3 \mathcal{S}(N)^{2 b} \left(4+\mathcal{S}(N)^{2 b}\right) \ln\left[12 \mathcal{S}(N)^{2 b}+3 \mathcal{S}(N)^{4 b}\right]\Big{)}\epsilon+\mathcal{O}(\epsilon^2)\, .
\end{align}
where the function $\mathcal{S}(N)$ stands for $\mathcal{S}(N)=-e^{N\alpha } G_0+G_1$. Accordingly, by using Eqs. (\ref{frgravity}), (\ref{hub2}), (\ref{cnc}) and (\ref{lagrange}), we obtain the Lagrange multiplier function $\lambda (N)$, which we quote in the Appendix. Below we present the approximate form of $\lambda (N)$, after we expanded in terms of the parameter $\epsilon$, which is,
\begin{align}\label{lajdbfe}
& \lambda (N)=\left(-3 \mathcal{S}(N)^{2 b}-\frac{3}{2} \mathcal{S}(N)^{4 b}-3 b \left(e^N\right)^{\alpha } G_0 \mathcal{S}(N)^{-1+2 b} \alpha \right)\\ \notag &
+\Big{(}3 \mathcal{S}(N)^{2 b}-\frac{1}{4+\mathcal{S}(N)^{2 b}}-3 b \left(e^N\right)^{\alpha } G_0 \mathcal{S}(N)^{-1+2 b} \alpha -3 \mathcal{S}(N)^{2 b} \ln\left[12 \mathcal{S}(N)^{2 b}+3 \mathcal{S}(N)^{4 b}\right]\\ \notag &
-\frac{3}{2} \mathcal{S}(N)^{4 b} \ln\left[12 \mathcal{S}(N)^{2 b}+3 \mathcal{S}(N)^{4 b}\right]
-3 b \left(e^N\right)^{\alpha } G_0 \mathcal{S}(N)^{-1+2 b} \alpha  \ln\left[12 \mathcal{S}(N)^{2 b}+3 \mathcal{S}(N)^{4 b}\right]\Big{)}\epsilon +\mathcal{O}(\epsilon^2)\, ,
\end{align}
with the function $\mathcal{S}(N)$ being as we defined it earlier. Therefore we demonstrated how the cosmology with Hubble rate as in Eq. (\ref{hub2}), may be realized by a mimetic $F(R)$ gravity with Lagrange multiplier, and as we shown, this cosmology is compatible with both Planck and BICEP2/Keck Array data. In principle, the presence of the Lagrange multiplier and of the mimetic potential in the theoretical framework, offers much freedom in producing viable cosmologies. Indeed, as we demonstrate in the next section, a different Hubble rate from that we used in Eq. (\ref{hub2}), can lead to a cosmological evolution which is also viable and compatible with both Planck and BICEP2/Keck Array observational data.

\subsection{Cosmological Model II}

In order to further support the utility of the mimetic $F(R)$ gravity formalism, we shall present another cosmological model, compatible with both Planck \cite{planck} and the BICEP2/Keck Array \cite{bicep2} data, which can be generalized from the mimetic $F(R)$ gravity. As we shall demonstrate for this case too, quite general cosmological evolutions can be realized due to the presence of the mimetic potential and of the Lagrange multiplier. Consider the following cosmological evolution,
\begin{equation}\label{hub1}
H(N)=\left(-G_0\text{  }N^{\alpha }+G_1\right)^b\, ,
\end{equation}
with the parameters $G_0$, $G_1$, $\alpha$ and $b$ being arbitrary real numbers. In order to avoid inconsistencies, which can occur if the Hubble rate of Eq. (\ref{hub1}) turns negative, we choose the parameter $b$ to be constrained as follows,
\begin{equation}\label{bparameter}
b=\frac{2n}{2m+1},\,\,\, b<1\, ,
\end{equation}
with $m$ and $n$ being arbitrary positive integers. By combining Eqs. (\ref{hub1}) and (\ref{hubbleslowrollnfolding}), the slow-roll parameter $\epsilon$, reads,
\begin{align}\label{hubpar1}
& \epsilon=N^{1-\alpha } \Big{(}G_1-G_0 N^{\alpha }\Big{)}\Big{(}3-\frac{b G_0
N^{-1+\alpha } \alpha }{G_1-G_0 N^{\alpha }}\Big{)}^{-2}(4 b G_0 \alpha )^{-1}
\Big{(}-\frac{6 b G_0 N^{-1+\alpha } \alpha }{G_1-G_0 N^{\alpha }}+\frac{b^2
G_0^2 N^{-2+2 \alpha } \alpha ^2}{\Big{(}G_1-G_0 N^{\alpha }\Big{)}^2} \\
\notag &
+\Big{(}G_1-G_0 N^{\alpha }\Big{)}^{-b} \Big{(}-b G_0 N^{-2+\alpha }
\Big{(}G_1-G_0 N^{\alpha }\Big{)}^{-1+b} (-1+\alpha ) \alpha  \\ \notag &
+(-1+b) b G_0^2 N^{-2+2 \alpha } \Big{(}G_1-G_0 N^{\alpha }\Big{)}^{-2+b}
\alpha ^2\Big{)}\Big{)}^2 \, .
\end{align}
and accordingly, the slow-roll parameter $\eta$ becomes equal to,
\begin{align}\label{etaparameterversion1}
& \eta=-\frac{G_1^2 (-1+\alpha ) (-3+6 N+\alpha )+G_0^2 N^{2 \alpha }
\left(3-10 b \alpha +8 b^2 \alpha ^2+6 N (-1+4 b \alpha )\right)}{4 N
\left(G_1-G_0 N^{\alpha }\right) \left(3 G_1 N-G_0 N^{\alpha } (3 N+b \alpha
)\right)}\\ \notag &
+\frac{-2 G_0 G_1 N^{\alpha } (3 N (-2+\alpha +4 b \alpha )+(-1+\alpha )
(-3+(-1+5 b) \alpha ))}{4 N \left(G_1-G_0 N^{\alpha }\right) \left(3 G_1 N-G_0
N^{\alpha } (3 N+b \alpha )\right)}\, .
\end{align}
The corresponding spectral index of primordial curvature perturbations $n_s$ easily follows by combining Eqs. (\ref{hubpar1}) and (\ref{etaparameterversion1}), and it is equal to,
\begin{align}\label{observa}
& n_s=\frac{1}{2 N \Big{(}G_1-G_0 N^{\alpha }\Big{)} \Big{(}-3 G_1 N+G_0
N^{\alpha } (3 N+b \alpha )\Big{)}^2}\times \\ \notag &
\Big{(}3 G_1^3 N \Big{(}-3+6 N (1+N)+4 \alpha -6 N \alpha -\alpha
^2\Big{)}-G_0^3 N^{3 \alpha } (9 N (-1+2 N (1+N)) \\ \notag &
 +48 b N^2 \alpha +2 b^2 (-1+13 N) \alpha ^2+4 b^3 \alpha ^3\Big{)}-G_0 G_1^2
N^{\alpha } \Big{(}54 N^3+2 b (-1+\alpha ) \alpha ^2+3 N (-1+\alpha ) (9+\alpha )
\\ \notag &  +6 N^2 (9-6 \alpha +8 b \alpha )\Big{)}+G_0^2 G_1 N^{2 \alpha }
\Big{(}54 N^3+2 b (1+b) (-1+\alpha ) \alpha ^2 \\ \notag &
 +6 N^2 (9-3 \alpha +16 b \alpha )+N \Big{(}-27+2 \alpha  \Big{(}6+3 \alpha +13
b^2 \alpha \Big{)}\Big{)}\Big{)}\Big{)} \, .
\end{align}
By choosing the free parameters $G_0$, $G_1$, $\alpha$
and $b$ in the following way,
\begin{equation}\label{defparam}
G_0=0.0000001,\, \, \, G_1=7900, \, \, \,\alpha=2.7, \, \, \, b=\frac{3}{2}\, ,
\end{equation}
and by substituting in the expressions of the observational indices, and for $N=50$ $e-$foldings, we acquire the following values for $n_s$ and $r$,
\begin{equation}\label{scalindex}
n_s\simeq 0.966034, \, \, \, r\simeq 8.72765 \times 10^{-7}\, .
\end{equation}
As it can be seen from Eq. (\ref{scalindex}), both $n_s$ and $r$ are compatible with both the Planck data \cite{planck} and the BICEP2/Keck Array data \cite{bicep2}. Let us demonstrate which mimetic potential and Lagrange multiplier generates the cosmological evolution (\ref{hub1}), for the same $F(R)$ gravity as before, given in Eq. (\ref{frgravity}). By combining Eqs. (\ref{hub1}), (\ref{frgravity}),
(\ref{cnc}), and (\ref{scalarpot}), we can straightforwardly calculate the mimetic scalar potential
$V(N)$, and we quote the result in the Appendix. Below we quote the mimetic potential expanded in powers of the parameter $\epsilon$, as $\epsilon\rightarrow 0$,
\begin{align}\label{explicitpot1}
& V(N)=\left(3 \mathcal{K}(N)^{2 b} \left(4+\mathcal{K}(N)^{2 b}\right)-2 \left(3 \mathcal{K}(N)^{2 b}-b G_0 N^{-1+\alpha } \mathcal{K}(N)^{-1+2 b} \alpha \right)\right) \\ \notag &
\Big{(} \frac{4}{3 \left(4+\mathcal{K}(N)^{2 b}\right)}+2 \left(\frac{4 b G_0 N^{-1+\alpha } \mathcal{K}(N)^{2 b} \left(2+\mathcal{K}(N)^{2 b}\right) \alpha }{\left(-G_1+G_0 N^{\alpha }\right) \left(4+\mathcal{K}(N)^{2 b}\right)}-\frac{1}{\left(-G_1+G_0 N^{\alpha }\right)^2 \left(4+\mathcal{K}(N)^{2 b}\right)^2}\right .\\ \notag &
4 b G_0 N^{-2+\alpha } \mathcal{K}(N)^{2 b} \alpha  \Big{(}-8 G_1+8 G_0 N^{\alpha }-6 G_1 \mathcal{K}(N)^{2 b}+6 G_0 N^{\alpha } \mathcal{K}(N)^{2 b}-G_1 \mathcal{K}(N)^{4 b}\\ \notag & +G_0 N^{\alpha } \mathcal{K}(N)^{4 b}+8 G_1 \alpha +6 G_1 \mathcal{K}(N)^{2 b} \alpha \Big{)}\\ \notag &
-\left.\left.4 b G_0 N^{\alpha } \mathcal{K}(N)^{2 b} \alpha +G_1 \mathcal{K}(N)^{4 b} \alpha \right)\right)+3 \mathcal{K}(N)^{2 b} \left(4+\mathcal{K}(N)^{2 b}\right) \ln\left[12 \mathcal{K}(N)^{2 b}+3 \mathcal{K}(N)^{4 b}\right]\Big{)}\epsilon+\mathcal{O}(\epsilon^2)\, ,
\end{align}
where we introduced the function $\mathcal{K}(N)=G_1-G_0 N^{\alpha }$, for notational simplicity. Accordingly, the full form of the Lagrange multiplier appears in the Appendix, and below we quote the $\epsilon$-expanded Lagrange multiplier, which reads,
\begin{align}\label{sder}
& \lambda (N)=\left(-3 \mathcal{K}(N)^{2 b}-\frac{3}{2} \mathcal{K}(N)^{4 b}-3 b G_0 N^{-1+\alpha } \mathcal{K}(N)^{-1+2 b} \alpha \right)\\ \notag &
+\Big{(}3 \mathcal{K}(N)^{2 b}-\frac{1}{4+\mathcal{K}(N)^{2 b}}-3 b G_0 N^{-1+\alpha } \mathcal{K}(N)^{-1+2 b} \alpha -3 \mathcal{K}(N)^{2 b} \ln\left[12 \mathcal{K}(N)^{2 b}+3 \mathcal{K}(N)^{4 b}\right]\\ \notag &
-\frac{3}{2} \mathcal{K}(N)^{4 b} \ln\left[12 \mathcal{K}(N)^{2 b}+3 \mathcal{K}(N)^{4 b}\right]-3 b G_0 N^{-1+\alpha } \mathcal{K}(N)^{-1+2 b} \alpha  \ln\left[12 \mathcal{K}(N)^{2 b}+3 \mathcal{K}(N)^{4 b}\right]\Big{)}\epsilon+\mathcal{O}(\epsilon^2)\, .
\end{align}
Hence, the cosmological evolution (\ref{hub1}) can be realized in the context of the mimetic $F(R)$ gravity with Lagrange multiplier, if the mimetic potential and the Lagrange multiplier are chosen as in Eqs. (\ref{explicitpot1}) and (\ref{sder}) respectively.

\section{Conclusions}

In this paper we explicitly demonstrated that by using the mimetic $F(R)$ formalism with Lagrange multiplier, it possible to achieve compatibility of the resulting theory with the recent BICEP2/Keck Array observational data. Actually in some cases, the predicted value of the scalar to tensor ratio $r$, can be well below the upper limit posed by the BICEP2/Keck Array observational data. In the formalism we employed, by sufficiently choosing a viable cosmological evolution and by fixing the $F(R)$ gravity, then the mimetic potential and the Lagrange multiplier can realize the specific chosen cosmology. One of the attributes of our formalism, is that for an arbitrarily chosen $F(R)$ gravity, in principle any viable cosmology can be realized by conveniently choosing the mimetic potential and the Lagrange multiplier. However, on of the drawbacks of our approach is the complexity of the functional form of the mimetic potential and of the Lagrange multiplier. In fact, the complexity of the mimetic potential and of the Lagrange multiplier increases as the complexity of the $F(R)$ gravity increases. However, the attribute of having a viable cosmological description compatible with observational data to a great extent, motivates us to further ask whether it can be possible to find a variant method which will yield simpler functional forms for the mimetic potential and Lagrange multiplier. In addition to this, in principle the $F(R)$ is arbitrarily chosen, but it must chosen in such a way so that it guarantees that inflation ends at some point. In fact, the graceful exit issue can be resolved in the mimetic $F(R)$ theories, as we demonstrated in \cite{mimeletter,mimeticbig}, if the theory contains unstable de Sitter vacua, which eventually are the final attractors of the cosmological dynamical system. In this case, the $F(R)$ gravity plays an important role in resolving the graceful exit problem of the theory. In conclusion, a refined version of the theory we used can effectively result into simpler forms for the mimetic potential and Lagrange multiplier and also effectively address the graceful exit issue that may occur if a simple $F(R)$ gravity is chosen.

\section*{Acknowledgments}

This work is supported in part by MINECO (Spain), project FIS2013-44881
(S.D.O) and partly by Min. of Education and Science of Russia (S.D.O and
V.K.O).

\section*{Appendix: The Explicit Forms of the Mimetic Potential $V(N)$ and of the Lagrange Multiplier $\lambda(N)$ Appearing in the Text}

In this Appendix we quote the exact forms of the mimetic potential $V(N)$ and of the Lagrange multiplier $\lambda (N)$, that can realize the cosmologies (\ref{hub2}) and (\ref{hub1}). We start off with the first cosmological evolution, namely (\ref{hub2}), for which, the exact form of the potential reads,
\begin{align}\label{vnmod1}
& V(N)=\left(12 \mathcal{S}(N)^{2 b}+3 \mathcal{S}(N)^{4 b}\right)^{1+\epsilon }+2 \mathcal{S}(N)^{2 b} \left(-3+\frac{b \left(e^N\right)^{\alpha } G_0 \alpha }{-\left(e^N\right)^{\alpha } G_0+G_1}\right)\\ \notag &
+4 \mathcal{S}(N)^{2 b} \left(12 \mathcal{S}(N)^{2 b}+3 \mathcal{S}(N)^{4 b}\right)^{-1+\epsilon } \epsilon  (1+\epsilon )\\ \notag &
\frac{1}{\left(4+\mathcal{S}(N)^{2 b}\right)^2}8\ 3^{\epsilon } b \left(e^N\right)^{\alpha } G_0 \mathcal{S}(N)^{2 (-1+b)} \left(\mathcal{S}(N)^{2 b} \left(4+\mathcal{S}(N)^{2 b}\right)\right)^{\epsilon } \alpha  \epsilon  (1+\epsilon )\\ \notag &
\left(-G_1 \left(8+6 \mathcal{S}(N)^{2 b}+\mathcal{S}(N)^{4 b}\right) (1+\alpha )+\left(e^N\right)^{\alpha } G_0 \left(8+6 \mathcal{S}(N)^{2 b}+\mathcal{S}(N)^{4 b}+4 b \alpha  \left(\mathcal{S}(N)^{2 b}+\left(2+\mathcal{S}(N)^{2 b}\right)^2 \epsilon \right)\right)\right)\, ,
\end{align}
while the corresponding Lagrange multiplier function $\lambda (N)$ reads,
\begin{align}\label{newlambda}
& \lambda (N)=-\frac{1}{2} \left(12 \mathcal{S}(N)^{2 b}+3 \mathcal{S}(N)^{4 b}\right)^{1+\epsilon }+3 \left(12 \mathcal{S}(N)^{2 b}+3 \mathcal{S}(N)^{4 b}\right)^{\epsilon } \left(\mathcal{S}(N)^{2 b}\right )\\ \notag &
\left.-b \left(e^N\right)^{\alpha } G_0 \mathcal{S}(N)^{-1+2 b} \alpha \right) (1+\epsilon )-3 \mathcal{S}(N)^{2 b} \left(12 \mathcal{S}(N)^{2 b}+3 \mathcal{S}(N)^{4 b}\right)^{-1+\epsilon } \epsilon  (1+\epsilon )
\end{align}
where recall that the function $\mathcal{S}(N)$ appearing in Eqs. (\ref{vnmod1}) and (\ref{newlambda}), is equal to, $\mathcal{S}(N)=\left(-e^N\alpha G_0+G_1\right)$. For the cosmological evolution (\ref{nvcompletemod2}), the exact form of the mimetic potential is equal to,
\begin{align}\label{nvcompletemod2}
& V(N)=\left(12 \mathcal{K}(N)^{2 b}+3 \mathcal{K}(N)^{4 b}\right)^{1+\epsilon }+\frac{\mathcal{K}(N)^{-1+2 b} \left(-6 G_1 N+2 G_0 N^{\alpha } (3 N+b \alpha )\right)}{N} \\ \notag &
+4 \mathcal{K}(N)^{2 b} \left(12 \mathcal{K}(N)^{2 b}+3 \mathcal{K}(N)^{4 b}\right)^{-1+\epsilon } \epsilon  (1+\epsilon )\\ \notag &
+\frac{1}{\mathcal{K}(N)^2 \left(4+\mathcal{K}(N)^{2 b}\right)^3}8\ 3^{\epsilon } b G_0 N^{-2+\alpha } \left(\mathcal{K}(N)^{2 b} \left(4+\mathcal{K}(N)^{2 b}\right)\right)^{1+\epsilon } \alpha  \epsilon  (1+\epsilon ) \left(-G_1 \left(8+6 \mathcal{K}(N)^{2 b}+\mathcal{K}(N)^{4 b}\right) (-1+N+\alpha )\right.\\ \notag & +\left.G_0 N^{\alpha } \left(-8-6 \mathcal{K}(N)^{2 b}-\mathcal{K}(N)^{4 b}+N \left(8+6 \mathcal{K}(N)^{2 b}+\mathcal{K}(N)^{4 b}\right)+4 b \alpha  \left(\mathcal{K}(N)^{2 b}+\left(2+\mathcal{K}(N)^{2 b}\right)^2 \epsilon \right)\right)\right)\, ,
\end{align}
while the Lagrange multiplier $\lambda (N)$ is equal to,
\begin{align}\label{lamdbaformodel2}
& \lambda (N)=\frac{1}{2} 3^{\epsilon } \left(\mathcal{K}(N)^{2 b} \left(4+\mathcal{K}(N)^{2 b}\right)\right)^{\epsilon } \Big{(}-12 \mathcal{K}(N)^{2 b}-3 \mathcal{K}(N)^{4 b}\\ \notag & +\frac{6 \mathcal{K}(N)^{-1+2 b} \left(G_1 N-G_0 N^{\alpha } (N+b \alpha )\right) (1+\epsilon )}{N}-\frac{2 \epsilon  (1+\epsilon )}{4+\mathcal{K}(N)^{2 b}}\Big{)}\, ,
\end{align}
where the function $\mathcal{K}(N)$ appearing in Eqs. (\ref{nvcompletemod2}) and (\ref{lamdbaformodel2}) was defined as $\mathcal{K}(N)=G_1-G_0 N^{\alpha }$.

\end{document}